\documentclass[%
 reprint,
superscriptaddress,
nofootinbib,
 amsmath,amssymb,
 aps,
]{revtex4-2}

\usepackage{graphicx}
\usepackage{dcolumn}
\usepackage{bm}
\usepackage{comment}
\usepackage{acronym}
\usepackage{color}

\usepackage[colorlinks]{hyperref}
\definecolor{LinkColor}{rgb}{0.75, 0, 0}
\definecolor{CiteColor}{rgb}{0, 0.5, 0.5}
\definecolor{UrlColor}{rgb}{0, 0, 0.75}
\hypersetup{linkcolor=LinkColor}
\hypersetup{citecolor=CiteColor}
\hypersetup{urlcolor=UrlColor}

\usepackage[dvipsnames]{xcolor}
\usepackage[normalem]{ulem}
\usepackage[capitalize]{cleveref}
\usepackage{xspace}
\usepackage{booktabs}
\usepackage{bookmark}

\definecolor{cadmiumgreen}{rgb}{0.0, 0.42, 0.24}

\newcommand{\dens}{\mathrm{~g~cm^{-3}}}

\newcommand{\num}{22\xspace}

\begin{document}

\preprint{APS/123-QED}

\title{Direct Measurement of the Accretion Disk Formed in Prompt Collapse Mergers with Future Gravitational-Wave Observatories}%

\newcommand{\psuigc}{\affiliation{Institute for Gravitation and the Cosmos, The Pennsylvania State University, University Park, PA, 16802, USA}}
\newcommand{\psuphys}{\affiliation{Department of Physics, The Pennsylvania State University, University Park, PA, 16802, USA}}
\newcommand{\psuastro}{\affiliation{Department of Astronomy \& Astrophysics, The Pennsylvania State University, University Park, PA, 16802, USA}}
\newcommand{\aei}{\affiliation{Max Planck Institute for Gravitational Physics (Albert Einstein Institute), Am Mühlenberg 1, Potsdam 14476, Germany}}
\newcommand{\infn}{\affiliation{Dipartimento di Fisica, Università di Trento, Via Sommarive 14, 38123 Trento, Italy}}
\newcommand{\trento}{\affiliation{INFN-TIFPA, Trento Institute for Fundamental Physics and Applications, Via Sommarive 14, I-38123 Trento, Italy}}
\newcommand{\UPisa}{\affiliation{Dipartimento di Fisica, Universit\`{a} di Pisa, Largo B.  Pontecorvo, 3 I-56127 Pisa, Italy}}
\newcommand{\infnPisa}{\affiliation{INFN, Sezione di Pisa, Largo B. Pontecorvo, 3 I-56127 Pisa, Italy}}
\newcommand{\gssi}{\affiliation{Gran Sasso Science Institute (GSSI), I-67100 L’Aquila, Italy}}
\newcommand{\infnassergi}{\affiliation{INFN, Laboratori Nazionali del Gran Sasso, I-67100 Assergi, Italy}}
\newcommand{\fbk}{\affiliation{Data Science for Industry and Physics, Fondazione Bruno Kessler, via Sommarive 18, 38123, Trento (TN), Italy}}

\author{Arnab Dhani}
\email{arnab.dhani@aei.mpg.de}
\aei

\author{Alessandro Camilletti}
\infn \trento \fbk

\author{Alessio Ludovico De Santis}
\gssi

\author{Andrea Cozzumbo}
\gssi

\author{David Radice}
\psuphys \psuigc \psuastro

\author{Domenico Logoteta}
\UPisa \infnPisa

\author{Albino Perego}
\infn \trento

\author{Jan Harms}
\gssi \infnassergi

\author{Marica Branchesi}
\gssi \infnassergi

\newcommand{\ad}[1]{\textcolor{teal}{{\it\textbf{[AD: #1]}}} }
\newcommand{\arnab}[1]{\textcolor{teal}{#1}}
\newcommand{\dr}[1]{\textcolor{purple}{{\it\textbf{[DR: #1]}}} }
\newcommand{\acozz}[2]{\textcolor{orange}{{\it\textbf{[ACozz: #1]}}} }
\newcommand{\ludo}[2]{\textcolor{Emerald}{{\it\textbf{[L: #1]}}} }

\begin{abstract}
The production site of heavy r-process elements, such as Gold and Uranium, is uncertain. Neutron star mergers are the only astrophysical phenomenon in which we have witnessed their formation~\cite{Perego:2021dpw,Kasen:2017sxr}. However, the amount of heavy elements resulting from the merger remains poorly constrained, mainly due to uncertainties on the mass and angular momentum of the disk formed in the merger remnant~\cite{Barnes:2020nfi}. Matter accretion from the disk is also thought to power gamma ray-bursts. We discover from numerical relativity simulations~\cite{Dhani:2023ijt} that the accretion disk influences the ringdown gravitational-wave signal produced by binaries that promptly collapse to black-hole at merger. We propose a method to \emph{directly} measure the mass of the accretion disk left during black hole formation in binary mergers using observatories such as the Einstein Telescope or Cosmic Explorer with a relative error of 10\% for binaries at a distance of up to 30~Mpc, corresponding to an event rate of 0.001 to 0.25 events per year.
\end{abstract}

\maketitle


\paragraph*{\label{sec:intro}Introduction.}
The collapse of a \ac{BNS} merger remnant is followed by a ``ringdown'' signal that is well described as a sum of damped sinusoidal oscillations at sufficiently late times, when \ac{BH} perturbation theory applies \cite{Dhani:2023ijt,Zhang:2020qlh}. This signal is produced as the newly formed, perturbed \ac{BH} relaxes to a stationary state via the emission of \acp{GW}, with characteristic complex frequencies called \acp{QNM}. For an isolated Kerr \ac{BH}, the \ac{QNM} frequencies depend only on the mass and angular momentum of the \ac{BH} and are independent of the source of the perturbation. 
However, if the \ac{BH} is surrounded by an astrophysical environment, such as an accretion disk, the \ac{QNM} spectra are modified and are no longer a property of the \ac{BH} alone.
Nevertheless, the response of the \ac{BH} continues to be described by damped sinusoids, albeit with modified complex frequencies~\cite{Nagar:2006eu,Barausse:2014tra}. 

The prompt collapse of a \ac{BNS} merger remnant generically leads to the formation of an accretion disk around the remnant \ac{BH}. While equal mass prompt-collapse \ac{NS} mergers have negligible matter outflows and the majority of nuclear matter from the disrupted \acp{NS} fall into the remnant \ac{BH}~\cite{Bauswein:2013yna}, unequal mass mergers form a substantive accretion disk around the remnant \ac{BH}~\cite{Bernuzzi:2020txg}. The presence of this accretion disk modifies the \ac{QNM} spectrum of the \ac{BH}, leaving an imprint on the \ac{GW} spectrum of the postmerger signal. This is also expected in the context of \ac{NS}-\ac{BH} binaries, where it was shown that the mass of the accretion disk and the fundamental ringdown frequency of the remnant \ac{BH} both depend on the compactness of the \ac{NS}~\cite{Tsao:2024tme}. 

The evolution of the accretion disk over the viscous/cooling timescale unbinds a significant fraction of its mass in the form of neutron-rich ejecta \cite{Chen:2006rra, Fernandez:2013tya, Perego:2014fma, Just:2014fka, Siegel:2017nub, Miller:2019dpt, Fujibayashi:2020jfr, Sprouse:2023cdm}. The latter are expected to produce heavy elements through the so-called r-process nucleosynthesis, whose radioactive decays power the kilonova transient \cite{Kasen:2017sxr, Metzger:2019zeh}. Moreover, the accretion of magnetized matter through the horizon of a spinning BH can produce the relativistic jet powering gamma-ray bursts \cite{Blandford:1977ds, Zhang:2003uk, Nakar:2007yr, Ciolfi:2018tal, Nakar:2019fza, Gottlieb:2023sja}. However, the efficiency at which the disk expels ejecta or the mass accretion rate is converted into jet energy are still very uncertain. For example, present simulations predict that the fraction of the disk that becomes unbound can be anywhere between 10 and 50\% \cite{Fernandez:2018kax}. Even when a kilonova is detected, the possibility to infer the mass of the ejecta from the electromagnetic signal is severely limited by several uncertainties, including the nuclear physics that governs the r-process nucleosynthesis and the atomic physics that shapes the kilonova emission \cite{Barnes:2020nfi}. A direct measure of the mass and angular momentum of the accretion disk will provide a valuable constraint directly entering the interpretation of kilonova and gamma-ray burst emissions in terms of the source properties.

Next-generation ground-based \ac{GW} detectors, such as \ac{CE}~\cite{Reitze:2019iox} and Einstein Telescope(ET)~\cite{Punturo:2010zz}, will be sensitive to signals up to 10 kHz~\cite{Gupta:2023lga,Branchesi:2023mws,Evans:2023euw,ET:2019dnz}. This presents a unique opportunity for directly observing the remnant and its associated astrophysical phenomena. Several studies in the literature have explored the scientific potential of observing \acp{GW} from the dynamics of the merger remnant prior to its collapse to \ac{BH}. This ranges from determining the kilonova and the associated r-process nucleosynthesis to independently determining the nuclear \ac{EoS}, to identifying phase transitions and the presence of exotic particles~\cite{Bauswein:2011tp, Hotokezaka:2012ze, Korobkin:2012uy, Hotokezaka:2011dh, Fernandez:2013tya, Bauswein:2013yna, Takami:2014zpa, Wanajo:2014wha, Vincent:2019kor, Bernuzzi:2015rla, Radice:2016rys, Kasen:2017sxr, Chatziioannou:2017ixj, Siegel:2017nub, Torres-Rivas:2018svp, Radice:2018pdn, Bauswein:2018bma, Most:2018eaw, Shibata:2019wef, Metzger:2019zeh, Perego:2021dpw, Breschi:2021xrx, Prakash:2023afe, Espino:2023mda, Sprouse:2023cdm}. In fact, one of the main themes of next-generation ground-based \ac{GW} detectors is to understand the physics of massive \ac{NS} remnants \cite{Evans:2021gyd,Bogdanov:2022faf,Abac:2025saz}. To the contrary, our focus here is the postmerger signal from the collapse of the merger remnant \cite{Dhani:2023ijt}. 

\begin{figure*}
    \centering
    \includegraphics[width=\columnwidth]{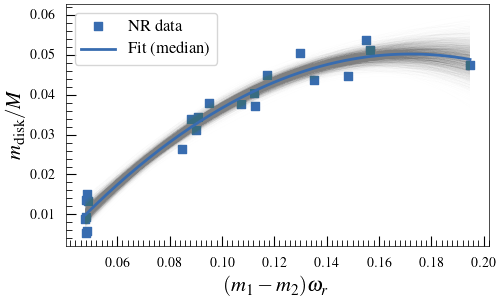}
    \includegraphics[width=\columnwidth]{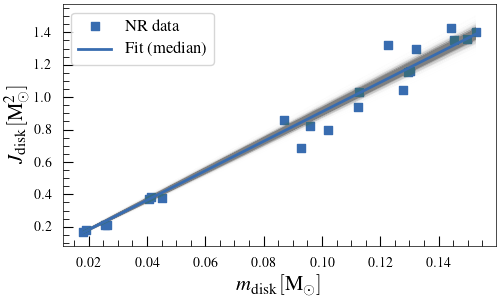}
    \caption{\textit{Left:} The quadratic relationship between the mass of the accretion disk and the \ac{QNM} oscillation frequency. \textit{Right:} The linear relationship between the disk's angular momentum and its mass. The markers show the values extracted from \ac{NR} simulations, while the blue curves show the fit for the median values of the parameters. The grey bands correspond to curves obtained by drawing from the fit parameters' posteriors and show the fit variance.}
    \label{fig:pymcfit}
\end{figure*}

In this \emph{Letter}, we use \num \ac{NR} simulations to show that the mass and angular momentum of the accretion disk produced in unequal mass prompt-collapse mergers can be measured using \acp{GW} from the perturbed remnant \ac{BH}. 
Detection of the postmerger signal with a \ac{SNR} of 5 can typically constrain the mass of the accretion disk to an accuracy of 10\%. 
Our work highlights the exciting potential of \ac{GW} science in the multi-kHz band.

\paragraph*{\label{sec:simulation}Simulations.}
We consider \num unequal mass \ac{BNS} simulations performed using the \texttt{WhiskyTHC} code~\cite{Radice:2012cu, Radice:2013xpa, Radice:2013hxh} starting from quasi-circular configurations created with \texttt{Lorene} \cite{Gourgoulhon:2000nn}. We use 4 \acp{EoS} spanning the current range of uncertainties, namely, BHB$\Lambda\phi$ \cite{Banik:2014qja}, HS(DD2) \cite{Typel:2009sy, Hempel:2009mc}, LS220 \cite{Lattimer:1991nc}, SFHo \cite{Steiner:2012rk} and consider total masses $M \in [2.8, 3.3]\rm\,M_{\odot}$ and mass ratios $q \in [0.6, 0.85]$. These simulations have been presented in Refs.~\cite{Perego:2021mkd, Camilletti:2022jms, Camilletti:2024otr}, to which we refer for a detailed description of the numerical setup.

\paragraph*{\label{sec:disk_properties}Disk properties.}
We extract the mass, $m_{\rm disk}$, and angular momentum, $J_{\rm disk}$, of the accretion disk surrounding the remnant \ac{BH} from our simulations by integrating the baryon rest mass density and the baryon angular momentum density profile outside the remnant \ac{BH}, respectively (see App.~\ref{sec:disk_extraction} for details). Given that the mass of the central object significantly exceeds the mass of the accretion disk, we neglect the minor contributions to the total disk mass arising from the kinetic energy of matter outflows and their associated gravitational potential energy \citep{Abramowicz:1984}. Moreover, to compute the angular momentum we assume that the system is axisymmetric about the orbital angular momentum of the binary. We find that this assumption is approximately satisfied following a relaxation phase after the merger of $10{-}15\ {\rm ms}$ \cite{Camilletti:2024otr}.

\paragraph*{\label{sec:model}Postmerger model.}
The \ac{GW} waveform morphology for a \ac{BNS} merger undergoing prompt collapse consists of a slowly increasing amplitude during the inspiral, followed by a rapid and steep dip indicating the merger, and a subsequent increase of the amplitude that finally decays exponentially. 
We build a phenomenological model of the most dominant quadrupolar radiation of the postmerger \ac{GW} signal following the dip in the amplitude (see \cref{sec:pc_morphology} for details). 
We construct the model so that the amplitude damps exponentially and the phase grows linearly at late times, describing a \ac{QNM} signal, while at early times, they describe the bump in the amplitude and the rapid growth of the phase. These ansatz are motivated by \ac{BBH} waveform models, which use analogous functional forms to transition from the inspiral to the ringdown. 

We find that the disk's mass correlates well with the oscillation frequency of the postmerger model. The left panel of Fig.~\ref{fig:pymcfit} shows the relation between $(m_1-m_2)\,\omega_r$ and $m_{\rm disk}/M$ where $m_1$ and $m_2$ are the component masses of the binary ($m_2<m_1$), and $M=m_1+m_2$ is the total mass. A quadratic function can model the relation well. We construct a simple Bayesian regression model given by $Y\sim\mathcal{N}(a+bX+cX^2,\sigma^2)$\footnote{We use $\mathcal{N}(\mu,\sigma^2)$ to denote a normal distribution with mean $\mu$ and standard deviation $\sigma$} where $Y=m_{\rm disk}/M$ and $X=(m_1-m_2)\omega_r$ and assume uninformative priors on all parameters, $\vec{\phi}=\{a,b,c,\sigma\}$. The blue line in Fig.~\ref{fig:pymcfit} shows the quadratic model with the median values of the parameters, while the thin gray curves show the probabilistic distribution of the model. 

The right panel of Fig.~\ref{fig:pymcfit} similarly shows the relation between the disk's mass and angular momentum, which we map using a linear function. We find that the 90\% highest density interval for the slope lies between $8.8-9.4$, consistent with previous works~\cite{Camilletti:2022jms,Camilletti:2024otr}.

\paragraph*{\label{sec:SNR}Postmerger SNR.}
We estimate the expected post-merger \acp{SNR} for \ac{BNS} mergers observed in proposed next-generation ground-based detectors, such as the \ac{ET} and \ac{CE}, using the \texttt{GWFish} code~\cite{Dupletsa:2022scg,Dupletsa:2024gfl}. The \ac{LWA}, typically used in GW signal analysis, holds at low frequencies, and, as a result, the detector transfer function is unity. However, the characteristic frequency at which this approximation breaks down is $f_c=c/2\pi L$, where $L$ is the detector arm length. Since our interest is in the kilo-Hertz frequency range, it is important to properly account for the detector transfer function, which deviates from unity at these frequencies \cite{Essick:2017wyl} (see \cref{sec:transfer_function} for details). This significantly degrades the sensitivity of these detectors at high frequencies. Additionally, note that the inverse relationship of $f_c$ to the detector arm length implies that the corrections are more important for longer detectors. This is why one can safely ignore these corrections for current detectors.

\begin{figure}[h]
    \centering
    \includegraphics[width=\columnwidth]{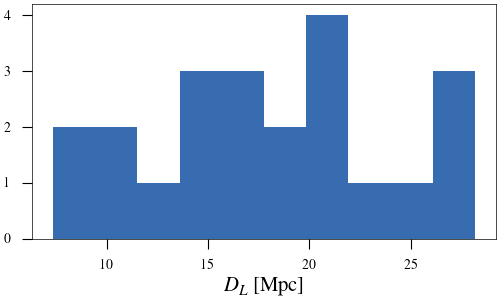}
    \caption{The distance at which the postmerger \ac{SNR} is 5 for the \num \ac{NR} simulations considered in this study}
    \label{fig:distance_snr5}
\end{figure}

We consider a three-detector network of future \ac{GW} observatories, consisting of a 40 km \ac{CE}, a 20 km \ac{CE}, and a triangular \ac{ET}. We place both the \ac{CE} detectors at the location of the LIGO-Hanford detector and the \ac{ET} detector at the location of the Virgo detector. The choice of the detector location is unimportant for our analysis and has little effect on our results. 
We show the distances at which the median --- calculated by uniformly distributing the sources across the plane of the sky, inclination angles of the orbit, and polarization angles --- post-merger \ac{SNR} is 5 in \cref{fig:distance_snr5}. Additionally, we report the median and 90\% interval for the \ac{SNR} distribution in \cref{tab:snr_results} of \cref{sec:transfer_function} for sources at a fixed distance of 40 Mpc.

\begin{figure}
    \centering
    \includegraphics[width=\columnwidth]{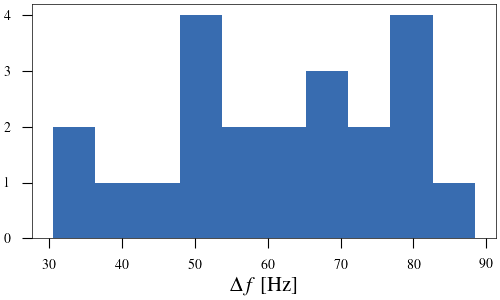}
    \caption{Distribution of the measurement errors in the oscillation frequency across the \num \ac{NR} simulations for a white noise \ac{SNR} of 5 in the ringdown signal.}
    \label{fig:osc_freq_err}
\end{figure}

\paragraph*{\label{sec:measurement_accuracies}Measurement prospects.}
A \ac{BNS} merger at the distances reported in \cref{fig:distance_snr5}
in future detectors will have an inspiral \ac{SNR} of $\mathcal{O}(10^3)$. This will allow the measurement of the component masses and the total mass of the binary from the inspiral with exquisite precision.
We assess the measurement error on the \ac{GW} observable $\omega_r$ 
from the post-merger signal
to estimate the accuracy of measuring the disk properties using \ac{GW} observations. 
To simplify our analysis, we focus on the case where $\omega$ is the only unknown parameter. This simplification is necessary because we do not have a model for the calibration parameters in relation to the binary parameters (see \cref{sec:pc_morphology}). However, it is a reasonable approximation since the primary quantity of interest is the oscillation frequency, which contains the most significant information at late times and exhibits minimal degeneracy with the overall amplitude scale. To obtain the measurement errors, we use a Fisher matrix approach, assuming white, Gaussian noise and an \ac{SNR} of 5 in the frequency range $3-10$ kHz. \Cref{fig:osc_freq_err} illustrates the distribution of the errors in $f=\omega_r/2\pi$ across the \num \ac{NR} simulations considered here, with a median error of $\sim60$ Hz. 
We recognize that achieving a post-merger \ac{SNR} of 5 is an optimistic case given the distances reported in \cref{fig:distance_snr5}, assuming current instrument design. We highlight the scientific potential of such an observation to encourage the development of technologies that improve the high-frequency sensitivity. Indeed, recent work~\cite{Jungkind:2025oqm} in this direction shows that the high-frequency sensitivity can be dramatically improved with novel interferometric designs.

\begin{figure}
    \centering
    \includegraphics[width=\columnwidth]{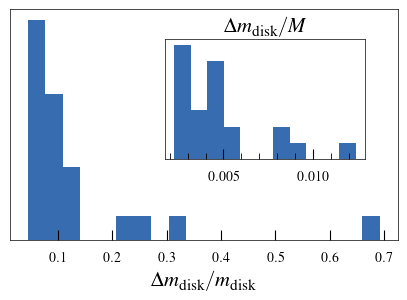}
    \caption{Distribution of the fractional errors on an accretion disk's mass for randomly sampled points across the domain of the fit in Fig.~\ref{fig:pymcfit}. The postmerger signal is assumed to have an \ac{SNR} of 5.}
    \label{fig:mdisk_hist}
\end{figure}

We can now predict the mass and angular momentum of the accretion disk using the fits in \cref{fig:pymcfit}. 
We illustrate the distribution of the fractional error on the disk mass in \cref{fig:mdisk_hist} assuming that $M$ is measured exactly.
We observe that the median measurement error is $\Delta m_{\rm disk}/m_{\rm disk}\lesssim 10\%$. Notice that a few measurements have a large fractional error, corresponding to the smallest values of $(m_1-m_2)\omega_r$. The absolute errors relative to the total mass of the binary $\Delta m_{\rm disk}/M$ are shown in the inset of \cref{fig:mdisk_hist}. We remark that the disk's mass can be measured exquisitely at better than $1\%$ of the total mass for most of the binaries we consider. Such a measurement will place stringent direct constraints on the accretion disk's mass.

We can further calculate the disk's angular momentum using the disk's mass measurements using the model on the right panel of \cref{fig:pymcfit}. We find that the uncertainty in the model is negligible and the typical measurement accuracy for the angular momentum of the disk is $\sim$10\%.

\paragraph*{\label{sec:discussion}Discussion.}
Current estimates of the rate of BNS mergers in the local Universe from the LIGO-Virgo-KAGRA Collaboration are highly uncertain. The rate of a \ac{BNS} merger similar to GW170817 originating at 40 Mpc is 1 in 12 years, assuming a fiducial rate of $300\,\rm Gpc^{-3}yr^{-1}$ consistent with~\textcite{KAGRA:2021duu}. On the other hand, we would expect 1 event within a distance of 100 Mpc every year. 
Furthermore, the mass distribution of astrophysical \acp{NS} in binaries is also highly uncertain, and therefore, the fraction of \ac{BNS} merger remnants promptly collapsing is currently unknown. 

\ac{NS} mergers are thought to be an important, if not dominant, site of production for r-process elements \cite{Perego:2021dpw}. Indeed, the kilonova AT2017gfo, associated with the \ac{NS} merger in GW170817, confirmed that \ac{NS} mergers produce r-process elements~\cite{Coulter:2017wya, Cowperthwaite:2017dyu, Tanvir:2017pws, Rosswog:2017sdn, Tanaka:2017qxj, Kasen:2017sxr, Perego:2017wtu, Watson:2019xjv}. However, the quantity and quality of the nucleosynthesis yields from mergers remain very uncertain, primarily due to the poorly known nuclear and atomic physics of neutron-rich nuclei~\cite{Barnes:2013wka, Tanaka:2013ana, Barnes:2016umi, Zhu:2018oay, Waxman:2019png, Tanaka:2019iqp, Barnes:2020nfi, Kato:2024jke, Mumpower:2024mlh}, which makes the interpretation of kilonova light curves and spectra challenging.

We demonstrate the possibility of a direct measurement of the mass and angular momentum of the accretion disk that forms around an unequal mass prompt-collapse neutron star merger using \ac{GW} observations. 
Knowing the properties of the accretion disk powering a kilonova -- even if for just one event -- would enable the community to calibrate and validate models used to interpret kilonova observations. This would be key to finally solve the puzzle of the origin of r-process elements, which has eluded the community for almost $70$ years~\cite{Burbidge:1957vc}.

A fraction of the accretion power of the disk onto the remnant \ac{BH} is used to power a collimated relativistic outflow thought to power \acp{GRB} through either the Blandford-Znajek mechanism \cite{Blandford:1977ds}, or neutrino-antineutrino annihilation \cite{1989Natur.340..126E, Zalamea:2010ax, Perego:2017fho}. In either case, the efficiency of this process depends strongly on the spin of the \ac{BH}, and, in the Blandford-Znajek case, on the amount of magnetic flux accumulated on the \ac{BH} horizon \cite{2011MNRAS.418L..79T, Gottlieb:2023sja}. The \ac{BH} mass and angular momentum can be obtained by subtracting the mass and angular momentum of the disk--measured following the approach outlined in this work--from the remnant mass and angular momentum, which is directly measurable through the \ac{GW} signal, e.g.,~\cite{Ruiz:2007yx}. As such, in combination with observations of the \ac{GRB} afterglow \cite{Paradijs:2000kh, Zhang:2005fa}, the direct detection of the ringdown signal from a promptly collapsing \ac{BNS} remnant would enable novel constraints on the mass accretion and horizon magnetic flux on the remnant \ac{BH}, and, more in general, provide a test for current theoretical models of \ac{GRB} engines.

Given the relevance of the response function at high frequency, we briefly discuss prominent \ac{CE} and \ac{ET} detector designs that are most suitable for studying \ac{BNS} merger remnants. 
With only the postmerger of massive \ac{NS} remnants in mind, studies in the literature, therefore, recommend a 20 km detector as optimal for \ac{NS} postmerger physics~\cite{Srivastava:2022slt,Branchesi:2023mws}. However, postmerger signals resulting from the collapse of the \ac{NS} remnant radiate at much higher frequencies, corresponding roughly to the \ac{QNM} frequencies of the resulting \ac{BH}~\cite{Dhani:2023ijt,Zhang:2020qlh}. 
Furthermore, for such signals, the full detector transfer function should be used for reliable estimates; studies often use the \ac{LWA}, which is not valid at these frequencies (see \cref{sec:transfer_function}). In fact, we encourage the consideration of a fuller range of \ac{BNS} postmerger physics while optimizing the design choices for future \ac{GW} observatories \cite{PhysRevD.64.042006,Ackley_2020}, as well as the development of new techniques to improve the sensitivity of such detectors at high frequency~\cite{Buonanno:2001cj,Ackley:2020atn,Jungkind:2025oqm}. 

We address some limitations of this work and suggest possible avenues for future research. Despite the use of 4 different EoS and the exploration of a broad range of mass ratios, our study relies on a limited set of available \ac{NR} simulations, which restricts the accuracy of the phenomenological relationship we have introduced. Nonetheless, due to the promising results presented here, we plan to conduct a larger series of high-resolution \ac{NR} simulations to explore the parameter space more thoroughly. Additionally, we developed a simple post-merger model to estimate the measurement accuracy of the accretion disk properties, as comprehensive post-merger models for promptly collapsing \ac{BNS} merger remnants are currently unavailable. Therefore, we defer a more detailed data analysis for the future, when such models can be utilized.

\begin{acknowledgments}
We thank Kevin Kuns and Mathew Evans for providing the \ac{CE} sensitivity curves up to 10 kHz and drawing our attention to the time-dependent nature of the antenna patterns in \ac{CE} at high frequencies.
DR acknowledges funding from the National Science Foundation under Grants No.~PHY-2020275, PHY-2116686, AST-2108467, PHY-2407681, the Sloan Foundation, and from the U.S. Department of Energy, Office of Science, Division of Nuclear Physics under Award Numbers DE-SC0021177 and DE-SC0024388.
MB and JH acknowledge the Astrophysics Centre for Multi-messenger studies in Europe (ACME) funded under the European Union’s Horizon Europe Research and Innovation programme under Grant Agreement No 101131928.
This research used resources of the National Energy Research Scientific Computing Center, a DOE Office of Science User Facility supported by the Office of Science of the U.S.~Department of Energy under Contract No.~DE-AC02-05CH11231.
\end{acknowledgments}

\bibliography{reference}
\clearpage

\section*{Supplementary materials}
\appendix

\section{\label{sec:pc_morphology}Waveform morphology}
\begin{figure}[h]
    \centering
    \includegraphics[width=\columnwidth]{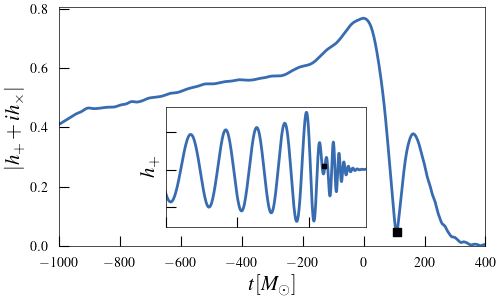}
    \caption{An illustrative waveform of a prompt collapse merger. The main plot shows the amplitude, while the inset shows the $+$ polarization. The black marker indicates the first minima of the amplitude following the peak and denotes the start of the postmerger model.}
    \label{fig:illustrative_waveform}
\end{figure}

In \cref{fig:illustrative_waveform}, we show the amplitude and plus polarization of one of the \ac{NR} simulations used in this study to illustrate the prominent features of a prompt collapse \ac{BNS} merger. The amplitude drops precipitously after reaching the peak before rising again and, finally, decaying exponentially. These generic features are present in all the \num \ac{NR} simulations that we consider. Our amplitude and phase model, given by the ansatz below, describes the signal after the dip, elucidated by the black marker in \cref{fig:illustrative_waveform}.

\begin{equation}
\begin{split}
    \mathcal{A}(t) &= (a_1 + a_2 \tanh(a_3 + a_4 t)) \exp(-\omega_i t), \\
    \Phi(t) &= \phi_0 + \omega_r t + b_1 \log(1 +  b_2 \exp(-b_3 t)).
    \label{eq:amp_phase_ansatz}
\end{split}
\end{equation}
Here, $a_i | i\in \{1,2,3,4\}$, $b_i | i\in \{1,2,3\}$, $\phi_0$, and $\omega \equiv \omega_r + i \omega_i \in \mathcal{C}$ are fit to \ac{NR} simulations with $\omega$ being the complex \ac{QNM} frequency. Though beyond the scope of this work, one can fit the parameters $a_i$ and $b_i$ across the binary parameter space to construct a semi-analytic model of the postmerger signal.

\section{\label{sec:ringdown_fit}Ringdown fit}
We fit the parameters in the amplitude and phase ansatzes in \cref{eq:amp_phase_ansatz} separately using non-linear least squares. The loss function that is minimized is the $L_2$-norm of the log-amplitude and the phase. The postmerger signal that we consider starts from the dip following the amplitude peak, as shown in \cref{fig:illustrative_waveform}, and extends till 1\% of the peak amplitude. We find that at later times, the noise dominates.

\section{\label{sec:disk_extraction}Extracting disk quantities}
The accretion disks in our simulations are torus-shaped dense clouds of matter surrounding the remnant \ac{BH}. Their maximum density reaches a few $10^{12} \dens$ near the interior edge of the disk.
We are interested in computing the mass, $m_{\rm disk}$, and the angular momentum, $J_{\rm disk}$, of the accretion disks in our simulations. To compute the latter quantities, we first remove from the computational domain the central \ac{BH}, considered as the region with a lapse function smaller than 0.2. The disk mass is then computed as the volume integral in cylindrical coordinates, $(r,\phi,z)$, of the rest mass density
\begin{equation}
\label{eq:disc_mass}
	m_{\rm disk} = \int \sqrt \gamma \rho W ~rdr d\phi dz ~,
\end{equation}
where $\rho$, $W$ and $\sqrt{\gamma}$ are the baryon rest mass density, the Lorentz factor of the fluid the determinant of the 3-metric, respectively.
We observe that, after a short relaxation phase occurring just after the merger, the space-time can be regarded as axisymmetric, i.e. $\partial_{\phi}$ is approximately a killing vector \cite{Radice:2018xqa}.
Assuming axial symmetry around the rotational axis of the binary, the angular momentum is computed as the volume integral of the baryon angular momentum density along the azimuthal direction,
\begin{equation}
\label{eq:disc_am}
	J_{\rm disk} = \int \sqrt \gamma \rho h W^2 \tilde v_{\phi} ~rdr d\phi dz ~,
\end{equation}
where $h$ is the fluid specific enthalpy and $\tilde v_{\phi}$ is the advective angular velocity in the azimuthal direction.

The properties of the disk change as a result of the complex dynamics that follows the prompt collapse of the central object. Processes such as accretion into the central BH, viscous effects, and neutrino absorption and emission contribute to the evolution of both the disk mass and angular momentum. We evaluate $m_{\rm disk}$ and $J_{\rm disk}$ at the last available time step of each numerical simulation.

\section{\label{sec:transfer_function}Transfer function and SNR computation}
\begin{figure}
    \centering
    \includegraphics[width=\linewidth]{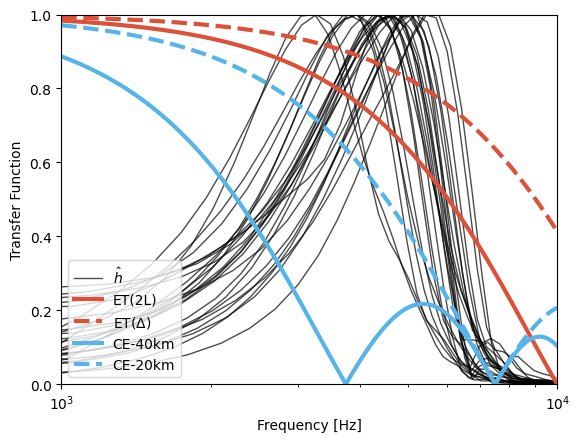}
    \caption{Transfer function for the considered interferometers. 
    Given that the transfer function depends on the incident angle of the GW wave, we only plot an incident angle of $ \theta_i = 0^o$. To give a qualitative impression on the impact of these transfer functions, we also plot all the 22 NS strains normalized with regards to their maxima ($\hat{h}$). While not considered in our cited results we also show the transfer function of the ET(2L) configuration for comparison.}
    \label{fig:transfer}
\end{figure}

The \ac{GW} strain measured at a detector is the projection of the independent \ac{GW} polarizations on to the detector transfer function. For a detector of arm length $L$ and a source direction $\hat{n}$ on the plane of the sky, it is defined as~\cite{Essick:2017wyl}
\begin{multline}
    D(f,\hat{n}) = \frac{e^{-2\pi i fL}}{2(1-\hat{n}^2)} \left( \frac{\sin(2\pi fL)}{2\pi fL} - \hat{n}^2 \frac{\sin(2\pi \hat{n} fL)}{2\pi \hat{n} fL} \right. \\ \left. - \frac{i\hat{n}}{2\pi fL} (\cos(2\pi fL) - \hat{n}^2 \cos(2\pi \hat{n} fL)) \right)
\end{multline}
In general, it is a function of both the direction of the incoming \ac{GW} and its frequency. However, its magnitude is unity in the \ac{LWA} as can be seen in \cref{fig:transfer} which shows the transfer function for different next-generation observatories. Notice that the function rapidly decreases from unity for higher frequencies adversely affecting the signal's \ac{SNR}. We report the postmerger \acp{SNR} in the realistic case and under the \ac{LWA} in \cref{tab:snr_results} for binaries placed at a fixed distance of 40 Mpc. 
We observe that the \ac{SNR} estimates decrease by $\sim$4 when relaxing the \ac{LWA}. This highlights the importance of optimizing the high-frequency performance of future \ac{GW} observatories to maximize the prospects of \ac{GW} science at high frequencies.

\begin{table}
\centering
\caption{The postmerger \ac{SNR} for the \num \ac{NR} simulations considered in this study in a detector network comprising a triangular \ac{ET} and two \acp{CE}, with each binary located at a distance of 40 Mpc. We quote the median and the 90\% range in the \ac{SNR} when a binary's position in the sky and its orientation with the observer are varied. We also show the impact of the \ac{LWA} at these high frequencies.}
\begin{tabular}{lcccc}
\toprule
EOS & $m_1$ ($M_\odot$) & $m_2$ ($M_\odot$) & SNR (No LWA) & SNR (LWA) \\
\midrule
BHB$\Lambda\phi$ & 1.775 & 1.065 & $0.92_{-0.36}^{+0.32}$ & $2.93_{-1.60}^{+2.56}$ \\
LS220 & 1.752 & 1.138 & $1.02_{-0.44}^{+0.51}$ & $3.47_{-2.00}^{+2.90}$ \\
SFHo & 1.719 & 1.031 & $1.24_{-0.55}^{+0.64}$ & $4.13_{-2.37}^{+3.64}$ \\
BHB$\Lambda\phi$ & 1.781 & 1.069 & $1.32_{-0.54}^{+0.56}$ & $4.08_{-2.41}^{+3.57}$ \\
LS220 & 1.704 & 1.278 & $1.51_{-0.62}^{+0.69}$ & $4.95_{-2.88}^{+4.42}$ \\
SFHo & 1.541 & 1.309 & $1.83_{-0.77}^{+0.76}$ & $5.75_{-3.33}^{+5.32}$ \\
SFHo & 1.535 & 1.305 & $1.91_{-0.77}^{+0.70}$ & $5.85_{-3.54}^{+4.94}$ \\
SFHo & 1.623 & 1.217 & $1.95_{-0.87}^{+0.92}$ & $6.44_{-3.66}^{+5.44}$ \\
LS220 & 1.720 & 1.204 & $2.00_{-0.81}^{+0.97}$ & $6.30_{-3.56}^{+5.57}$ \\
LS220 & 1.697 & 1.272 & $2.01_{-0.87}^{+0.86}$ & $6.48_{-3.80}^{+5.31}$ \\
LS220 & 1.691 & 1.269 & $2.09_{-0.89}^{+0.88}$ & $6.85_{-4.03}^{+6.12}$ \\
BHB$\Lambda\phi$ & 1.649 & 1.401 & $2.30_{-1.07}^{+1.26}$ & $8.29_{-4.76}^{+6.90}$ \\
LS220 & 1.605 & 1.365 & $2.35_{-0.93}^{+1.12}$ & $7.75_{-4.70}^{+6.75}$ \\
HS(DD2) & 1.935 & 1.355 & $2.52_{-1.09}^{+1.16}$ & $8.64_{-5.38}^{+7.48}$ \\
HS(DD2) & 1.957 & 1.273 & $2.61_{-1.18}^{+1.43}$ & $8.79_{-4.89}^{+8.14}$ \\
BHB$\Lambda\phi$ & 1.747 & 1.223 & $2.68_{-1.18}^{+1.36}$ & $9.40_{-5.81}^{+9.02}$ \\
HS(DD2) & 1.784 & 1.516 & $2.69_{-1.25}^{+1.23}$ & $8.63_{-5.15}^{+7.31}$ \\
HS(DD2) & 1.929 & 1.351 & $2.78_{-1.11}^{+1.33}$ & $9.72_{-5.75}^{+8.55}$ \\
HS(DD2) & 1.778 & 1.512 & $3.00_{-1.30}^{+1.69}$ & $10.13_{-6.12}^{+8.01}$ \\
HS(DD2) & 1.880 & 1.410 & $3.37_{-1.40}^{+1.60}$ & $11.14_{-6.57}^{+9.61}$ \\
BHB$\Lambda\phi$ & 1.643 & 1.397 & $3.38_{-1.44}^{+1.55}$ & $11.74_{-7.04}^{+9.75}$ \\
BHB$\Lambda\phi$ & 1.714 & 1.286 & $3.52_{-1.48}^{+1.70}$ & $12.08_{-7.44}^{+10.52}$ \\
\bottomrule
\end{tabular}
\label{tab:snr_results}
\end{table}

\acrodef{BNS}{binary neutron star}
\acrodef{NS}{neutron star}
\acrodef{GW}{gravitational wave}
\acrodef{NR}{numerical relativity}
\acrodef{BH}{black hole}
\acrodef{QNM}{quasi-normal mode}
\acrodef{EM}{electromagnetic}
\acrodef{EoS}{equation of state}
\acrodefplural{EoS}[EoS]{equations of state}
\acrodef{CE}{Cosmic Explorer}
\acrodef{ET}{Einstein Telescope}
\acrodef{SNR}{signal-to-noise ratio}
\acrodef{BBH}{binary black holes}
\acrodef{EOB}{Effective one-body}
\acrodef{LWA}{long-wavelength approximation}
\acrodef{GRB}{gamma-ray burst}

\end{document}